# Electrically-controlled digital metasurface device for light projection displays

*Jianxiong Li[1], Ping Yu[1], Shuang Zhang[2], and Na Liu[3,4*]*

[1]Max Planck Institute for Intelligent Systems, Heisenbergstrasse 3, 70569 Stuttgart, Germany
[2]School of Physics & Astronomy, University of Birmingham, Birmingham B15 2TT, UK.
[3]Kirchhoff Institute for Physics and Centre for Advanced Materials, University of Heidelberg, Im Neuenheimer Feld 227, 69120 Heidelberg, Germany.
[4]Max Planck Institute for Solid State Research, Heisenbergstrasse 1, 70569 Stuttgart, Germany
*e-mail: na.liu@kip.uni-heidelberg.de

Light projection displays play an increasingly important role in our modern life. Core projection systems including liquid crystal displays and digital micromirror devices can impose spatial light modulation and actively shape light waves. Recently, the advent of metasurfaces has revolutionized the design concepts in nanophotonics, enabling a new family of optical elements with exceptional degrees of freedom. Here, we demonstrate a light projection display technology based on optical metasurfaces, called digital metasurface device (DMSD). Each metasurface pixel in a DMSD is electrically reconfigurable with well-controlled programmability and addressability. The DMSD can not only continuously modulate the intensity of light with high contrast, but also shape the wavefront of light generated by each metasurface pixel and dynamically switch between arbitrary holographic patterns. Our approach will pave an avenue towards the development of customized light projection devices. It will also drive the field of dynamic optical metasurfaces with fresh momentum and inspiring concepts.

The fast development of nanophotonics urges breakthrough concepts and design principles to create functional devices for diverse optical applications. In particular, the advent of metasurfaces has revolutionized nanophotonics, enabling a variety of unprecedented opportunities. Metasurfaces, which comprise subwavelength optical antennas in ultrathin layers, identify a new class of flat optical devices with exceptional control over the propagation of light [1]. Over the past years, many functional metasurfaces have been demonstrated, ranging from beam deflectors [2-4], wave plates [5,6], flat lenses [7-9], displays [10,11], holograms [12-14] to surface wave couplers [15,16] and among others.

Despite the exciting progress, a specific research focus - dynamic optical metasurfaces [17], which probably also provide the most important link to the real-world optical applications, still awaits endeavors. They allow for active manipulation of light beams and enable fascinating optical functions, including dynamic beam steering, focusing and shaping, as well as optical vortex generations, imaging, and sensing. Many schemes have been explored to dynamically control the metasurfaces, for instance, by mechanical [18,19], chemical [11,12], electrical [20-25], thermal [26,27], and magnetic [28] external stimuli. However, the reported schemes have been mostly limited to simultaneous tuning of the resonance frequencies or amplitudes of all the metasurface antennas. They also suffer from poor device control with low intensity modulation and narrow operating band. Thus, there is still plenty of room to advance this research focus, especially at visible frequencies.

In this work, we demonstrate a novel strategy to realize dynamic optical metasurfaces by tailoring their spatial frequencies via modulation of both the geometric and propagation phases at visible frequencies. In particular, we showcase electrically-controlled digital metasurface devices (DMSDs) for light projection displays. Figure 1 illustrates the general concept. The DMSD consists of metasurface pixels in an M ×N array. Each metasurface pixel contains gold nanorods arranged in a rectangular lattice. In some preselected (odd or even) columns, the nanorods are covered with a dielectric material. The sample is subsequently encapsulated in a thin liquid crystal (LC) cell. The dynamic function of the metasurface pixels is enabled by electrically controlling the relative phase between the neighboring odd and even columns via LCs on the millisecond time scale. In principle, each metasurface pixel can be designed to generate a specific dynamic holographic pattern in the far field. Particularly for light projection display applications, each metasurface pixel that can be electrically switched between '1' and '0' states is individually addressed by an electrode, forming an anomalous reflection spot that can be turned on and off with high contrast in the projection screen. As a result, programmable images are dynamically generated and displayed as illustrated in Fig. 1.

To illustrate the working principle, we first describe a specific optical device, comprising $i \times j$ elements, with a discrete phase distribution profile to generate anomalous reflection in the $y$ direction as follows:

$$\varphi_{ij} = \begin{cases} \varphi_j, & i = 1, 3, \cdots, 2n+1 \\ \varphi_j + \Delta\varphi, & i = 2, 4, \cdots, 2n \end{cases}, \qquad (1)$$

$$\varphi_j = \frac{\pi}{2}(j-1), \quad j = 1, 2, \cdots n,$$

where $\varphi_{ij}$ represents the phase delay from the element positioned at $(i, j)$ and $n$ is an integer. Thus, every element pair, formed by the two elements located in the neighboring odd and even columns differ by a phase factor of $\Delta\varphi$. A linear geometric phase gradient is also introduced along each column by spatially varying the orientation angle of the antennas. The detailed information can be found in Supplementary Fig. 1a. We take the spacing between the neighboring rows and columns both as 300 nm and the operating wavelength is 633 nm. After the discrete Fourier transform for such a phase profile, two frequency components can be obtained as shown in Fig. 2a. The red dot with the low frequency component $(v_x, v_y)/\lambda^{-1} = (0, 0.528)$ is located in the grey zone of the spatial frequency spectrum. It represents anomalous reflection of light with an angle of $\theta_y = \sin^{-1} \lambda v_y$ based on the generalized Snell's law. Meanwhile, the blue dot, with the high frequency component $(v_x, v_y)/\lambda^{-1} = (1.055, 0.528)$, corresponds to the excitation of evanescent waves (see Fig. 2a).

The phase factor $\Delta\varphi$ can be utilized to readily achieve the transformation between the output propagating and evanescent waves. Figure 2b presents the calculated intensities of the low and high spatial frequency components in dependence on $\Delta\varphi$. With increase of $\Delta\varphi$, the anomalous reflection first diminishes in intensity and is gradually transformed into an evanescent wave. When $\Delta\varphi$ reaches $\pi$, it is completely turned off due to the destructive interference between the contributions from neighboring columns. With further increase of $\Delta\varphi$, the reverse transformation takes place. The anomalous reflection therefore emerges again and is completely turned on, when $\Delta\varphi = 2\pi$.

It is noteworthy that the phase factor $\Delta\varphi$ is a generalized parameter applied in the design of optical metasurfaces. It may contain contributions from the geometric phase ($\Delta\varphi_g$) of an optical antenna determined by its orientation, and/or the propagation phase ($\Delta\varphi_p$) introduced by a dielectric material of a certain thickness covering the optical antenna (see Supplementary Fig. 2). If only the Pancharatnam-Berry (PB) phase $\Delta\varphi_g$ is adopted, then $\Delta\varphi = \Delta\varphi_g = 2\Delta\theta$. Here $\Delta\theta$ defines the angle difference between the gold nanorods located in the neighboring odd and even columns (see the inset of Fig. 2c and Supplementary Fig. 1b). As shown by the experimental and simulated results in Fig. 2c, the intensity of the anomalously reflected light from the metasurface can be modulated by simply tuning $\Delta\varphi$. In particular, when $\Delta\varphi = 2\Delta\theta = \pi$, the anomalous reflection completely vanishes.

If both $\Delta\varphi_g$ and $\Delta\varphi_p$ are applied (*i.e.*, $\Delta\varphi = \Delta\varphi_g + \Delta\varphi_p$), extra degrees of tunability to the metasurface can be achieved. As shown in Fig. 2d, the gold nanorods on a SiO$_2$ (100 nm)/gold mirror substrate are first embedded in a uniform spacer (green) with a refractive index of 1.5 and thickness of 50 nm. The alternating columns on the metasurface are then covered by two dielectric materials with refractive indices of $n_a$ (blue) and $n_b$ (pink), respectively (see Fig. 2d). When $\Delta\theta = \frac{\pi}{2}$, $\Delta\varphi_g$ is fixed at $\pi$. By tuning $n_a$ and $n_b$, $\Delta\varphi_p$ is altered accordingly. As a result, the intensity of the anomalously reflected light generated by the metasurface can be continuously modulated as demonstrated in Fig. 2e. Especially, it completely vanishes, when $n_a = n_b$ (see Supplementary Fig. 3).

Our concept can be employed to build dynamic metasurfaces for a broad range of optical functions. In particular, we demonstrate a DMSD for light projection displays. Figure 3a shows the schematic of the DMSD. An array of gold nanorods spaced by a dielectric polymer (PC403, 100 nm) resides on a gold electrode. The gold nanorods are then embedded in another layer of PC403 (50 nm) in order to eliminate the influence on the LC alignment (also see Supplementary Fig. 4). The alternating columns are covered by high-birefringence LCs ($n_a$) and PMMA ($n_b$) trenches of thickness $t$, respectively. Therefore, $n_b$ is fixed in this case. $\Delta\varphi_p$ can be finely tuned, when $n_a$ is dynamically changed upon electric control of the LCs. To reduce ion migration, the LC cell is driven by a 1 kHz AC sine wave. The incident light (633 nm) is linearly polarized along the PMMA trenches (see Supplementary Fig. 5). There are four addressable metasurface pixels (M1-M4) fabricated on the device, which are controlled via four independent electrodes as shown by the scanning electron microscopy (SEM) image in Fig. 3b. A polyimide alignment layer is rubbed along the direction of the PMMA trenches. The LC cell is encapsulated by an indium tin oxide (ITO)-coated quartz superstrate as a shared electrode.

When the cell is switched off, the long-axis of the LC molecules are aligned along the PMMA trenches. This corresponds to the case of $n_a = n_e = 1.92$ and $n_b = n_{PMMA} = 1.5$. By choosing an appropriate PMMA thickness, which is $t = 240$ nm in this case, one gets $\Delta\varphi = \Delta\varphi_g + \Delta\varphi_p = 2\pi$. This gives rise to the '1' state of the metasurface pixel and a bright anomalous reflection spot is generated in the far field. When the cell is switched on and the applied voltage (V) increases, the orientation of the LCs changes in response to the external electric field (see Fig. 3a). $n_a$ decreases gradually to $n_o$ = 1.53, which is approximately equal to $n_b$ = $n_{PMMA}$ = 1.5. As a result, $\Delta\varphi = \Delta\varphi_g + \Delta\varphi_p = \pi$. This leads to the '0' state of the metasurface pixel and thus the anomalous reflection spot disappears in the far field. With four independent metasurfaces, each representing an addressable pixel, a 4-bit DMSD is achieved. Programmable optical information ('1111', '0111', *etc.*) displayed as dynamic images on the project screen are generated by electrical control as shown in Fig. 3c (see also Supplementary Movie 1).

Figure 3d presents the intensity of the anomalously reflected light as a function of the applied voltage. It is evident that the light intensity continuously decreases, when the applied voltage increases. This results from the gradually changing $n_a$ of the LCs through electric control, while $n_b$ of PMMA is fixed. The experimental observation agrees nicely with the theoretical prediction presented in Fig. 2e. Notably, an intensity modulation as high as 105:1 is experimentally achieved. The switching time is 40 and 65 milliseconds for the '0' and '1' states, respectively (see Supplementary Fig. 6). Also, there is no evident signal degradation up to 100 cycles, demonstrating excellent reversibility (see Fig. 3e). It is also worth mentioning that only the LC molecules residing in the PMMA trenches with a height of $t$ (~ $\lambda/3$) contribute to the modulation of $\Delta\varphi_p$. Therefore, the DMSD thickness can be in principle subwavelength. This is in direct contrast to the working principle of liquid crystal displays that require much thicker LC layers. Importantly, the switching rate of the DMSD can be further enhanced by optimizing the structural parameters and reducing the LC thickness.

To further expand the capability of our DMSDs, we demonstrate a numeric indicator display, composed of seven electrically-controlled holographic segments. Figure 4a shows the photograph of the device. There are seven addressable metasurface pixels (M1-M7) on the device as presented by the SEM image in Fig. 4b. An enlarged view of one of the metasurface pixels is shown in Supplementary Fig. 7a. Each metasurface pixel is utilized to generate and switch on/off a holographic pattern, which corresponds to a segment of the numeric indicator as illustrated by the inset image in Fig. 4b. The off-axis angle information for the individual holographic patterns is indicated in Supplementary Fig. 7b. The optical setup for characterizing the device is presented in Fig. 4c. The reconstructed holographic images are projected onto a screen and captured by a visible camera. Figure 4d shows the experimental results, in which numbers from 0-9 (0 is not shown), corresponding to 7-bit optical information are dynamically displayed using one single device (see also Supplementary Movie 2).

Due to the fact that the DMSDs can not only modulate the light intensity but also the light wavefront generated from each metasurface pixel, arbitrary holographic images can be dynamically reconstructed. A representative example is shown in Fig. 5. The metasurface is multiplexed by two sets of gold nanorods located in subunit cells I and II, respectively, to generate two independent phase profiles based on Gerchberg-Saxton algorithm (see Fig. 5a). This gives rise to two off-axis holographic images of a traffic light man, representing 'stop' and 'walk' signs, respectively. In both I and II, the two neighboring nanorods embedded in PC403 are placed orthogonal to each other. One of the nanorods in I or II is further covered by PMMA ($n_{PMMA}$ = 1.5) or hafnium dioxide ($n_{HfO2}$ = 1.94, see also Supplementary Fig. 8). The SEM image of the metasurface is presented in

Fig. 5b. When the applied voltage increases from 0 to 30 V, the refractive index ($n_a$) of the covered LCs is varied from $n_e$ = 1.92 to $n_o$ = 1.53. At V = 0 V, subunit cell I is on ($n_e \neq n_{PMMA}$), whereas II is off ($n_e \approx n_{HfO2}$). The 'stop' sign appears as shown in Fig. 5c (see also Supplementary Movie 3). In contrast, At V = 30 V, subunit cell I is off ($n_o \approx n_{PMMA}$), whereas II is on ($n_o \neq n_{HfO2}$). As a result, the 'walk' sign shows up. At an intermediate voltage, for instance V = 5 V, both signs are visible in the far field.

In conclusion, we have demonstrated electrically-controlled DMSDs by dynamically tailoring the spatial frequencies of the metasurfaces. Such optical devices comprise independently tunable metasurface pixels, which can be readily extended to a large number with each occupying a much smaller area. The DMSDs for light projection display applications exhibit remarkable performance with high intensity contrast, fast switching rate on the millisecond time scale, and excellent reversibility. Importantly, our concept is universal. It works not only with LCs but also with a variety of active materials that exhibit refractive index changes upon electrical, light, thermal, or other external stimuli. Such active media include spiropyran molecules in response to light, synthesized polymers in response to pH changes, vanadium dioxide and germanium antimony telluride in response to temperature tuning, *etc*. Our work will largely enrich the functionality breadth of current metasurface devices at visible frequencies, advancing the field dynamically forward.

**References**

1   Yu, N. & Capasso, F. Flat optics with designer metasurfaces. *Nat. Mater.* **13**, 139-150 (2014).

2   Yu, N. *et al.* Light propagation with phase discontinuities: generalized laws of reflection and refraction. *Science* **334**, 333-337 (2011).

3   Huang, L. *et al.* Dispersionless phase discontinuities for controlling light propagation. *Nano Lett.* **12**, 5750-5755 (2012).

4   Wang, K. *et al.* Quantum metasurface for multiphoton interference and state reconstruction. *Science* **361**, 1104-1108 (2018).

5   Yu, N. *et al.* A broadband, background-free quarter-wave plate based on plasmonic metasurfaces. *Nano Lett.* **12**, 6328-6333 (2012).

6   Wu, P. C. *et al.* Versatile Polarization Generation with an Aluminum Plasmonic Metasurface. *Nano Lett.* **17**, 445-452 (2017).

7   Rubin, N. A. *et al.* Matrix Fourier optics enables a compact full-Stokes polarization camera. *Science* **365**, eaax1839 (2019).

8   Wang, S. *et al.* A broadband achromatic metalens in the visible. *Nat. Nanotech.* **13**, 227-232 (2018).

9   Lin, R. J. *et al.* Achromatic metalens array for full-colour light-field imaging. *Nat. Nanotech.* **14**, 227-231 (2019).

10  Zhu, X., Vannahme, C., Hojlund-Nielsen, E., Mortensen, N. A. & Kristensen, A. Plasmonic colour laser printing. *Nat. Nanotech.* **11**, 325-329 (2016).

11  Duan, X., Kamin, S. & Liu, N. Dynamic plasmonic colour display. *Nat. Commun.* **8**, 14606 (2017).

12   Li, J. *et al.* Addressable metasurfaces for dynamic holography and optical information encryption. *Sci. Adv.* **4**, eaar6768 (2018).

13   Zheng, G. *et al.* Metasurface holograms reaching 80% efficiency. *Nat. Nanotech.* **10**, 308-312 (2015).

14   Wen, D. *et al.* Helicity multiplexed broadband metasurface holograms. *Nat. Commun.* **6**, 8241 (2015).

15   Huang, L. *et al.* Helicity dependent directional surface plasmon polariton excitation using a metasurface with interfacial phase discontinuity. *Light Sci. Appl.* **2**, e70 (2013).

16   Sun, S. *et al.* Gradient-index meta-surfaces as a bridge linking propagating waves and surface waves. *Nat. Mater.* **11**, 426 (2012).

17   Shaltout, A. M., Shalaev, V. M. & Brongersma, M. L. Spatiotemporal light control with active metasurfaces. *Science* **364** (2019).

18   Holsteen, A. L., Raza, S., Fan, P., Kik, P. G. & Brongersma, M. L. Purcell effect for active tuning of light scattering from semiconductor optical antennas. *Science* **358**, 1407-1410 (2017).

19   Holsteen, A. L., Cihan, A. F. & Brongersma, M. L. Temporal color mixing and dynamic beam shaping with silicon metasurfaces. *Science* **365**, 257-260 (2019).

20   Zeng, B. *et al.* Hybrid graphene metasurfaces for high-speed mid-infrared light modulation and single-pixel imaging. *Light Sci. Appl.* **7**, 51 (2018).

21   Bartholomew, R., Williams, C., Khan, A., Bowman, R. & Wilkinson, T. Plasmonic nanohole electrodes for active color tunable liquid crystal transmissive pixels. *Opt. Lett.* **42**, 2810-2813 (2017).

22  Decker, M. *et al.* Electro-optical switching by liquid-crystal controlled metasurfaces. *Optics Express* **21**, 8879-8885 (2013).

23  Buchnev, O., Ou, J. Y., Kaczmarek, M., Zheludev, N. I. & Fedotov, V. A. Electro-optical control in a plasmonic metamaterial hybridised with a liquid-crystal cell. *Optics Express* **21**, 1633-1638 (2013).

24  Li, S.-Q. *et al.* Phase-only transmissive spatial light modulator based on tunable dielectric metasurface. *Science* **364**, 1087-1090 (2019).

25  Wu, P. C. *et al.* Dynamic beam steering with all-dielectric electro-optic III–V multiple-quantum-well metasurfaces. *Nat. Commun.* **10**, 3654 (2019).

26  Yin, X. *et al.* Beam switching and bifocal zoom lensing using active plasmonic metasurfaces. *Light Sci. Appl.* **6**, e17016 (2017).

27  Choi, C. *et al.* Metasurface with Nanostructured $Ge_2Sb_2Te_5$ as a Platform for Broadband-Operating Wavefront Switch. *Adv. Opt. Mater.* **7**, 1900171 (2019).

28  Zubritskaya, I., Maccaferri, N., Inchausti Ezeiza, X., Vavassori, P. & Dmitriev, A. Magnetic Control of the Chiroptical Plasmonic Surfaces. *Nano Lett* **18**, 302-307 (2018).

**Acknowledgements**

**General:** We thank the 4[th] Physics Institute at the University of Stuttgart for kind permission to use their electron-gun evaporation system. We thank Dr. Y. B Guo for help with the sample processing as well as Prof. N. Fruehauf and Dr. P. Schalberger for helpful discussions.

**Funding:** This project was supported by the European Research Council (ERC Dynamic Nano and ERC Topological) grants.

**Author contributions:** J.X.L. and N.L. conceived the project. P.Y. and S. Z. provided crucial suggestions to the main concept of the project. J.X.L. performed the experiments and theoretical calculations. All authors discussed the results, analyzed the data, and commented on the manuscript.

**Competing interests:** The authors declare no competing financial interest.

**Data and materials availability:** The data that support the plots within this paper and other findings of this study are available from the corresponding author upon reasonable request.

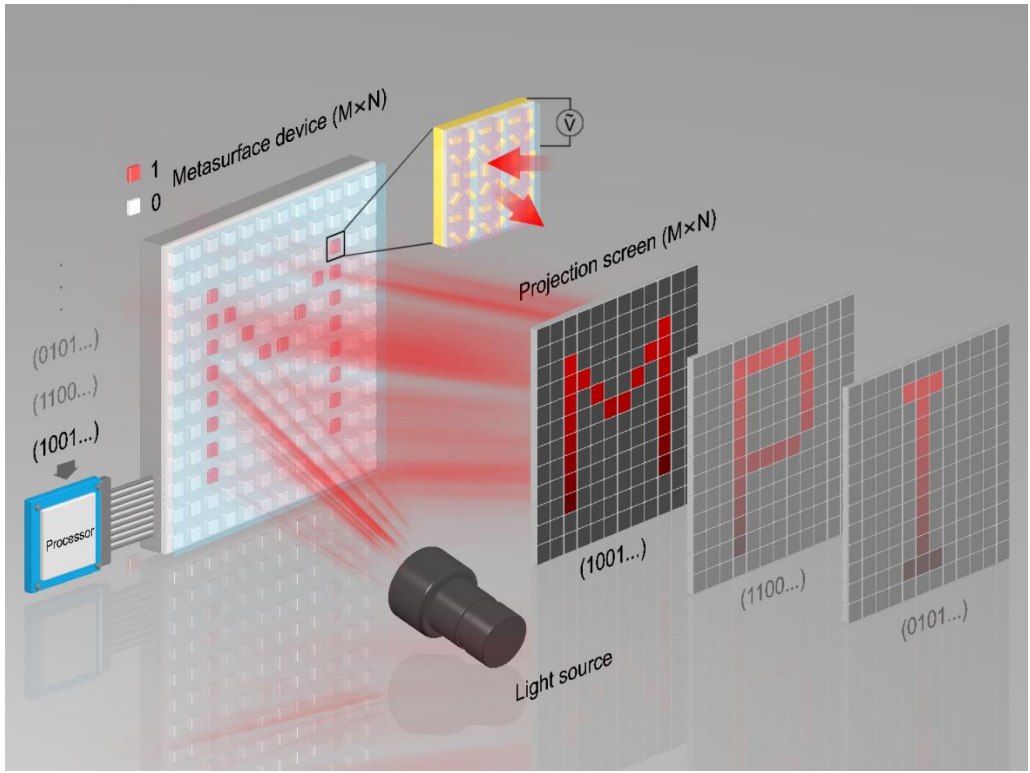

**Figure 1| Schematic of the electrically-controlled digital metasurface device (DMSD) for light projection displays.** The DMSD comprises metasurface pixels in an M × N array. Each metasurface pixel contains gold nanorods arranged in a rectangular lattice. The preselected columns are covered with a dielectric material. The sample is encapsulated in a thin liquid crystal (LC) cell. The intensity of the anomalous reflection from each metasurface pixel is controlled by an addressable electrode and can be independently switched on and off. Programmable images (M×N resolution) are dynamically displayed in the far field.

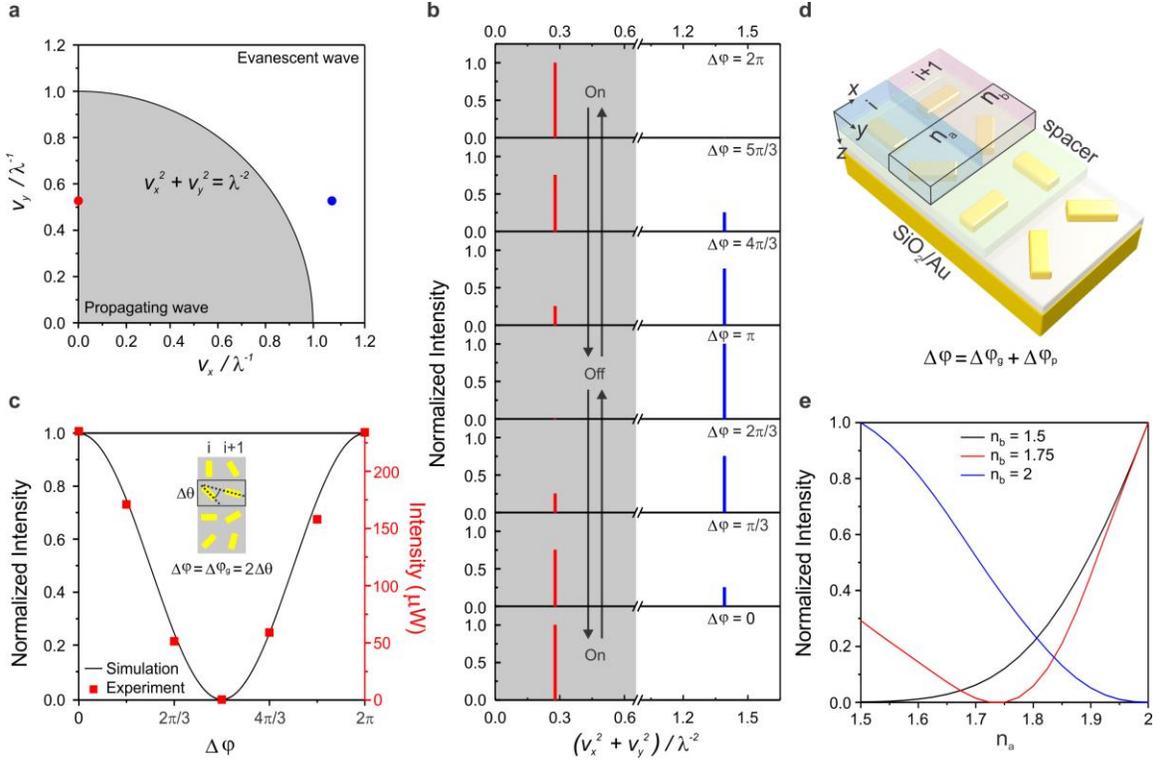

**Figure 2| Mechanism of dynamic spatial frequency modulation. a**, Two-dimensional spatial frequency spectrum calculated from the discrete phase distribution profile defined in Eq. (1). The spacing between the optical elements in the neighboring rows and columns are both taken as 300 nm and the operating wavelength is 633 nm. The circle of $v_x^2 + v_y^2 = \lambda^{-2}$ separates the zones of output propagating (grey) and evanescent (white) waves. Red and blue dots correspond to the low and high frequency components, respectively. **b**, Calculated intensities of the low and high spatial frequency components in dependence on $\Delta\varphi$. The zones of the propagating and evanescent waves are indicated in grey and white, respectively. **c**, Experimental and simulated results of the anomalous reflection intensity modulated by $\Delta\varphi$. $\Delta\theta$ defines the angle difference between the gold nanrods located in the neighboring odd and even columns, respectively. Each gold nanrod has a dimension of 200 nm × 80 nm × 30 nm. They reside on a gold mirror spaced by a SiO$_2$ layer (100 nm). **d**, Schematic of the reflective metasurface, in which the gold nanorods are embedded in a spacer with a refractive index of 1.5 and thickness of 50 nm. The alternating columns are further covered by two dielectric materials with refractive indices of $n_a$ and $n_b$, respectively. $\Delta\varphi$ contains contributions from both geometric ($\Delta\varphi_g$) and propagation phases ($\Delta\varphi_p$), *i.e.*, $\Delta\varphi = \Delta\varphi_g + \Delta\varphi_p$. **e**, Simulated intensity of the anomalously reflected light in dependence on $n_a$ and $n_b$ tuning.

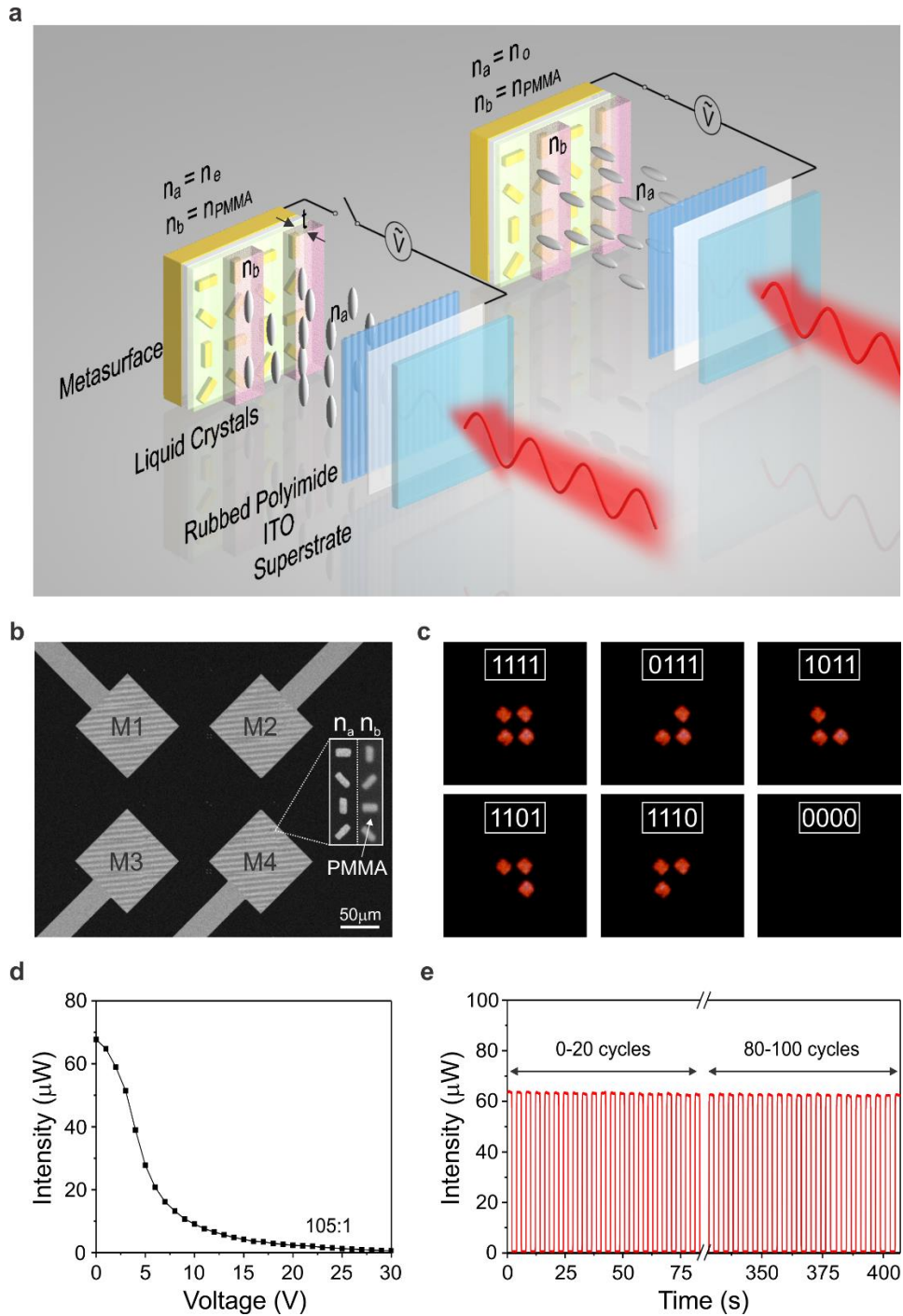

**Figure 3| Working principle and performance of the 4-bit DMSD. a**, Schematic of the electrically-controlled DMSD. An array of gold nanorods spaced by PC403 resides on a gold electrode. All nanorods are embedded in another PC403 layer of 50 nm. The alternating columns are then either covered by high-birefringence LCs ($n_a$) or by a PMMA ($n_b$) trench of thickness $t$. $\Delta\varphi_p$ is finely tuned, when $n_a$ is dynamically changed upon electric control of the LCs and $n_b$ is fixed. $n_o$ and $n_e$ are the ordinary and extraordinary refractive indices of the LCs, respectively. The incident light (633 nm) is linearly polarized

along the PMMA trenches. **b**, SEM image of the DMSD. Four addressable metasurface pixels (M1-M4) are controlled via four independent gold electrodes. Inset: enlarged SEM image of the gold nanorods in two neighboring columns covered by $n_a$ and $n_b$, respectively. **c**, Programmable optical information ('1111', '0111', *etc*.) is dynamically generated by the 4-bit DMSD. **d**, Intensity of the anomalously reflected light as a function of the applied voltage V. An intensity modulation ratio as large as 105:1 is achieved. **e**, Cycling performance of the DMSD, demonstrating excellent reversibility.

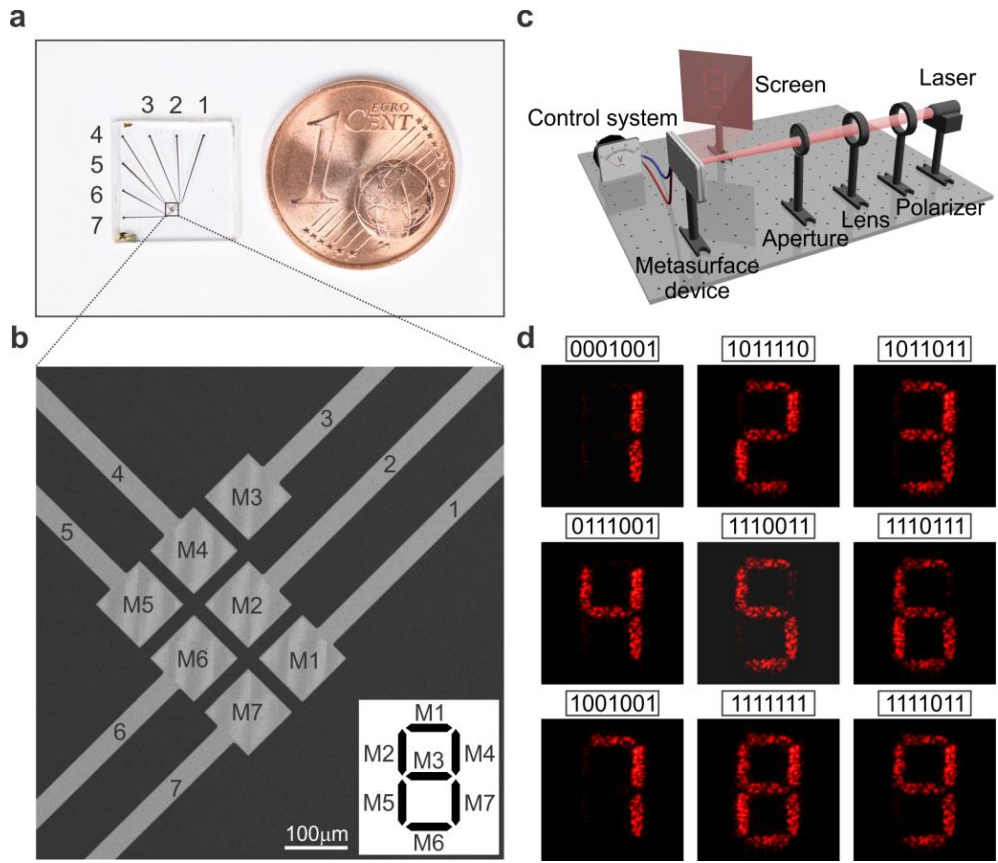

**Figure 4| Numeric indicator display composed of 7-segement holographic patterns generated by the electrically-controlled DMSD. a**, Photograph of the DMSD next to a 1 cent euro coin. **b**, SEM image of the DMSD with seven metasurface pixels, independently controlled by addressable electrodes. Inset: relation between each metasurface pixel and the corresponding holographic pattern for the numeric indicator display. **c**, Schematic of the optical setup. **d**, Experimental results. Different numbers corresponding to 7-bit optical information are dynamically generated by the DMSD.

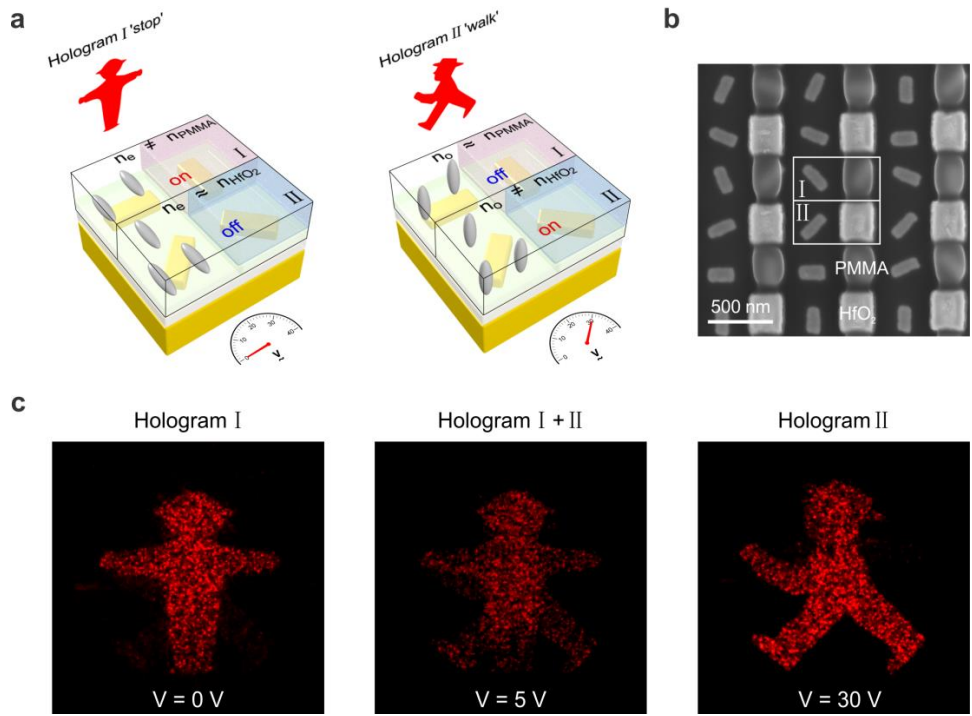

**Figure 5| Electrically-controlled dynamic holograms. a**, Schematic of the electrically-controlled metasurface device for switchable holograms between 'stop' and 'walk' signs. **b**, SEM image of the metasurface with selective PMMA and $HfO_2$ coating. **c**, Experimental results. Captured holographic images in the far field at different applied voltages.